\documentclass[11pt,letterpaper,pdf]{article}
\usepackage{jheppub}

\newcommand{\grtsim}{\mbox{\raisebox{-4pt}{$\stackrel{>}{\sim}$}}}
\newcommand{\lessim}{\mbox{\raisebox{-4pt}{$\stackrel{<}{\sim}$}}}

\title{
Bottom Quark Mass from $\Upsilon$ Sum Rules to  ${\cal O}(\alpha_s^3)$}

\author[a,b,c]{Alexander A. Penin}
\author[a]{and Nikolai Zerf}

\affiliation[a]{Department of Physics, University of Alberta,\\
  Edmonton AB T6G 2J1, Canada}
\affiliation[b]{Institut f{\"u}r Theoretische Teilchenphysik, Karlsruhe
  Institute of Technology (KIT),\\
  76128 Karlsruhe, Germany}
\affiliation[c]{Institute for Nuclear Research, Russian Academy of Sciences,\\
  119899 Moscow, Russia}

\emailAdd{penin@ualberta.ca}
\emailAdd{zerf@ualberta.ca}

\abstract{We use the  ${\cal O}(\alpha_s^3)$ approximation of the heavy-quark
vacuum polarization function in the threshold region to determine the bottom
quark mass from nonrelativistic $\Upsilon$ sum rules. We find very good
stability and convergence of the perturbative  series for the bottom quark mass
in $\overline{\rm MS}$ renormalization scheme. Our final result is
$\overline{m}_b(\overline{m}_b)=4.169\pm 0.008_{th}\pm 0.002_{\alpha_s}\pm
0.002_{exp}$.
}

\newcommand{\bfm}[1]{\mbox{\boldmath$#1$}}

\begin{document}

\hspace*{110mm}\mbox{ALBERTA-THY-02-14}
\vspace*{-10.5mm}

\maketitle
\section{Introduction}
The bottom quark mass $m_b$ is a fundamental parameter of the Standard Model of
particle interactions.  It is an essential input  parameter  for the analysis of
the  $B$-meson decays and CKM quark mixing matrix  as well as the Higgs boson
decay rates and branching ratios.  The precise value of the bottom quark mass is
also crucial for testing possible extensions of  the Standard Model such as
grand unified theories. Thus determination of the bottom quark mass with the
best possible precision is an important problem of particle phenomenology. A
unique  tool for such a determination is given by the analysis of the family of
$\Upsilon$ resonances within quantum chromodynamics. Direct application of
perturbative QCD  to the  description of the heavy-quarkonium properties such as
 the resonance  mass and width suffers from sizable long-distance
nonperturbative effects \cite{Voloshin:1978hc,Leutwyler:1980tn} resulting in
large uncertainties even if high-order approximation is available
\cite{Brambilla:1999xj,Penin:2002zv,Kniehl:1999mx,Kniehl:2002yv,Beneke:2014qea}.
The sum  rules approach \cite{Novikov:1976tn,Novikov:1977dq} suggests an elegant
solution of the problem. It relates the moments of the spectral density
saturated with the contribution of  $\Upsilon$ resonances to the derivatives of
the heavy quark vacuum polarization  function in a deep Euclidean region, which
can be reliably computed in perturbation theory. Thus the approach  provides a
model independent determination of the bottom quark mass entirely based on the
first principles of QCD. The low-moment or ``relativistic'' and the high-moment
or ``nonrelativistic'' sum rules require essentially different experimental and
theoretical input and can be considered as complimentary methods. Both methods
have been extensively  applied to the bottom quark mass determination. The most
recent analysis of the low-moment sum rules includes the third-order corrections
in the strong coupling constant $\alpha_s$ to the leading-order result
\cite{Kuhn:2007vp,Chetyrkin:2009fv,Chetyrkin:2010ic}.  At the same time the
high-moment  sum rules have been evaluated only through the
next-to-next-to-leading order (NNLO)
\cite{Voloshin:1987rp,Voloshin:1995sf,Kuhn:1998uy,Penin:1998zh,Hoang:1998uv,
Melnikov:1998ug,Penin:1998kx,Hoang:1999ye,Beneke:1999fe} though the effect of
higher order logarithmically enhanced terms have been considered
\cite{Pineda:2006gx,Hoang:2012us}. In this paper we present the  complete ${\cal
O}(\alpha_s^3)$ corrections to the heavy quarkonium  parameters required for the
N$^3$LO analysis of  the nonrelativistic  $\Upsilon$ sum rules and apply the
result to the determination of the bottom quark mass. In the next section we
outline the main concept of the nonrelativistic $\Upsilon$ sum rules and
describe  the perturbative approximation. The numerical analysis is given in
Section~\ref{sec::numerics}. In Section~\ref{sec::summary} our new estimate of
the bottom quark mass is compared to the existing high-order results.
\section{$\Upsilon$ sum rules}
\label{sec::sumrules}
We consider the vacuum polarization function $\Pi(q^2)$ defined through the
two-point vacuum correlator of the  heavy-quark electromagnetic current
$j_\mu=\bar b\gamma_\mu b$
\begin{equation}
\left(q_\mu q_\nu-g_{\mu \nu}q^2\right)\Pi(q^2)=
i\int d^dx\,e^{iqx}\,\langle 0|Tj_{\mu}(x)j_{\nu}(0)|0\rangle \,.
\label{eq::vacuumpol}
\end{equation}
Its $n$th moment is given by the normalized derivative
\begin{equation}
{\cal M}_n=
\left.{12\pi^2\over n!}(4m_b^2)^n{d^n\over ds^n}\Pi(s)\right|_{s=0}=
(4m_b^2)^n\int_0^\infty{R(s)ds\over s^{n+1}} \,,
\label{eq::momentdef}
\end{equation}
where $s=q^2$ and  $R(s)=12\pi {\rm Im}\Pi(s+i\epsilon)$ is the spectral
density. The long-distance nonperturbative contribution to
Eq.~(\ref{eq::momentdef}) is parametrically suppressed as
$\Lambda_{QCD}^4/m_b^4$ and the moments can be reliably computed in perturbative
QCD \cite{Novikov:1977dq}. On the other hand the optical theorem  relates the
spectral density to the experimentally measured cross section of $b\bar b$
hadron production in electron-positron annihilation
\begin{equation}
R_{exp}(s) = {1\over Q_b^2}{\sigma(e^+e^-\rightarrow {b\bar b})\over
\sigma(e^+e^-\rightarrow \mu^+\mu^-)} \,,
\label{eq::spectralexp}
\end{equation}
where $Q_b=-1/3$ is the bottom quark electric charge.
The moments of experimental spectral density~(\ref{eq::spectralexp}) get
dominant contribution from the $\Upsilon(nS)$ resonances
\begin{equation}
{\cal M}^{exp}_n={(4m_b^2)^n}{9\pi\over  Q_b^2\alpha^2(2m_b)}\left(
\sum_{m}{\Gamma_{\Upsilon(mS)\to l^+l^-}\over M_{\Upsilon(mS)}^{2n+1}}
+\ldots\right)\,,
\label{eq::momentexp}
\end{equation}
where $\alpha(2m_b)$ is the running QED coupling constant,
$M_{\Upsilon(mS)}$ ($\Gamma_{\Upsilon(mS)\to l^+l^-}$) is the resonance mass
(leptonic width), and ellipsis stand for the nonresonant contribution. The
sum rules then read
\begin{equation}
{\cal M}^{exp}_n={\cal M}^{th}_n\,,
\label{eq::sumrules}
\end{equation}
where the theoretical moment ${\cal M}^{th}_n$ is evaluated with the
perturbative QCD approximation for the spectral density. For small $n$ the
moments get sizable contribution from relativistic region above $b\bar b$ pair
threshold where the uncertainty of the measured cross section is relatively
large. For large $n$ the experimental moments are saturated with the
contribution of the lowest   $\Upsilon$ resonances measured with very high
accuracy. Moreover for large $n$  the moments~(\ref{eq::momentexp})  are very
sensitive to the value of  $m_b$, which significantly reduces the uncertainty of
the extracted bottom quark mass. The number of a phenomenologically relevant
moment is limited only by the magnitude of nonperturbative contribution which
grows as $n^3$ and becomes sizable for $n\grtsim 20$ \cite{Voloshin:1995sf}. The
perturbative description of the high moments, however, is nontrivial. For large
$n$ the theoretical moments are saturated with the nonrelativistic threshold
region where the heavy quark velocity $v$ is of order $1/\sqrt{n}$
\cite{Voloshin:1987rp}. This results in enhancement of the Coulomb effects which
are characterized by the expansion parameter $\alpha_s/v\sim\sqrt{n}\alpha_s$
rather then $\alpha_s$. For $n\grtsim 1/\alpha_s^2\sim 10$ the Coulomb terms are
not suppressed and have to be resummed to all orders. Thus for high moments the
QCD perturbation theory should be build up about the  nonrelativistic Coulomb
solution instead of the free heavy-quark approximation used for analysis of the
low moments.   In the next section we outline  how this can be done
systematically within the nonrelativistic effective field  theory framework.

\subsection{Effective theory approach to nonrelativistic sum rules}
\label{sec::quark}
In the threshold region the heavy-quark velocity $v$ is a small parameter. An
expansion in $v$ may be performed directly in the QCD Lagrangian by using the
concept of effective field theory \cite{Caswell:1985ui}.  The relevant modes are
characterized by the hard  ($k^0,{\bfm k} \sim m_q $), the soft ($k^0,{\bfm k}
\sim m_q v$), the potential ($k^0 \sim m_qv^2$, ${\bfm k}\sim m_q v $), and the
ultrasoft ($k^0,{\bfm k} \sim m_q v^2$) scaling of energy $k^0$ and
three-momentum $\bfm k$ in respect to the heavy-quark mass $m_q$.  Integrating
out the hard modes matches QCD onto non-relativistic QCD (NRQCD)
\cite{Bodwin:1994jh}.  By subsequent integrating out the soft modes and
potential gluons one obtains the  effective theory of potential NRQCD (pNRQCD)
\cite{Pineda:1997bj,Brambilla:1999xf,Kniehl:1999ud,Beneke:1998jj}, which
contains potential heavy quarks and ultrasoft gluons as dynamical fields
relevant for the description of the  heavy-quark threshold dynamics. In this
theory the propagation of a color-singlet quark-antiquark pair is described by
the Green function $G^{s}({\bfm r},{\bfm r}';E)$. In the leading-order Coulomb
approximation the Green function satisfies the Schr{\"o}dinger equation
\begin{equation}
\left({\cal H}_C-E\right)G_C^{s}({\bfm r},{\bfm r}';E)
=\delta^{(3)}({\bfm r}-{\bfm r}')\,,
\label{eq::schroedinger}
\end{equation}
with the  Hamiltonian
\begin{eqnarray}
{\cal H}_C&=& -{\bfm \partial^2\over m_q}-\frac{\alpha_s C_F}{r}\,,
\label{eq::hamilton}
\end{eqnarray}
where $r=|{\bfm r}|$, $m_q$ the heavy-quark pole mass, and
$C_F=(N_c^2-1)/(2N_c)$, $N_c=3$. The  leading-order Green function gets
corrections  due to the  high-order terms in the nonrelativistic and
perturbative expansion of the effective Hamiltonian  as well as due to the
multipole interaction of the potential  quarks to the ultrasoft gluons. In the
effective theory the electromagnetic  current is represented  by a series of
operators composed of the nonrelativistic quark and antiquark two-component
Pauli spinor fields $\psi$ and $\chi$
\begin{equation}
{\bfm j}=c_v\psi^\dagger{\bfm \sigma}\chi+{d_v\over6 m_q^2}
\psi^\dagger{\bfm \sigma}\mbox{\bfm D}^2\chi
+\ldots\,,
\label{qe::currentnr}
\end{equation}
where the matching coefficients $c_v=1+{\cal O}(\alpha_s)$ and $d_v=1+{\cal
O}(\alpha_s)$ are the series  in $\alpha_s$. The threshold behaviour of the
heavy-quark polarization function is given by the following pNRQCD expression
\begin{equation}
\Pi(q^2)={N_c\over 2 m_q^2}\left(c_v-{E\over m_q}{d_v\over 6}+\ldots\right)^2
\left(1+{E\over 2m_q}\right)^{-2} G^{s}(0,0;E)\,,
\label{eq::vacuumpolnr}
\end{equation}
where $E=\sqrt{q^2}-2m_q\sim v^2m_q$ and only the component of total spin one is
kept in the Green function as dictated by the  form of the production current.
The corrections to the leading order Coulomb approximation for the polarization
function~(\ref{eq::vacuumpolnr}) can be computed in pNRQCD as a series in
$\alpha_s$ and  $v\sim\alpha_s$ according to  the effective theory power
counting, which gives a systematic perturbative expansion for the high moments.
The explicit result for the polarization function up to the NNLO can be found in
Refs.~\cite{Penin:1998kx,Melnikov:1998ug}.  The spectral representation of the
color-singlet Green function includes an infinite number of Coulomb-like bound
state poles
\begin{equation}
G^s(0,0;E)=\sum_{n=1}^\infty{|\psi_n(0)|^2\over
  E_n-E-i\epsilon} +\ldots\,,
\label{eq::spectralrep}
\end{equation}
where the ellipsis stand for the continuum contribution, and $E_n$
($\psi_n({\bfm r})$) is the  energy (wave function) of the bound state with spin
$S=1$ and orbital angular momentum $l=0$. In the Coulomb approximation they read
\begin{equation}
E_n^{C} = -{m_q C_F^2\alpha_s^2\over 4n^2},
\qquad   |\psi^C_n(0)|^2={(m_q C_F \alpha_s)^3\over 8\pi n^3}\,.
\label{eq::coulomb}
\end{equation}
Thus, the effective theory expression for the  moments can be written in the
following form
\begin{equation}
{\cal M}_n=(4m_b^2)^n\left(
{12\pi^2N_c\over m_b^2}\sum_{m}{ Z_m\over (2m_b+E_m)^{2n+1}}
+\int_{4m_b^2}^\infty{R(s)ds\over s^{n+1}}\right)\,,
\label{eq::momentnr}
\end{equation}
where
\begin{equation}
Z_m= \left(c_v-{E_m\over m_b}{d_v\over 6}\right)^2
\left(1+{E_m\over 2m_b}\right)^{-2}{|\psi_m(0)|^2}\,,
\label{eq::zdef}
\end{equation}
and $R(s)$ for $s>4m_b^2$ is determined by the imaginary part of
Eq.~(\ref{eq::vacuumpolnr}). The quantities $E_n$ and $Z_n$ determine the
perturbative QCD predictions for the $\Upsilon(nS)$ resonance mass and leptonic
width
\begin{equation}
M_{\Upsilon(nS)}^{p.t.}=2m_b+E_n\,, \qquad
\Gamma_{\Upsilon(nS)\to l^+l^-}^{p.t.}=
{4\pi\over 3}{N_c Q_b^2\alpha^2(2m_b)\over m_b^2}Z_n\,.
\label{eq::upsilonmasswidth}
\end{equation}
The nonperturbative corrections to the binding energy and the width in
Eq.~(\ref{eq::upsilonmasswidth}) are parametrically as large as
$\Lambda^4_{QCD}/(\alpha_s^6m_b^4)$, being significantly enhanced in comparison
to the moments \cite{Voloshin:1978hc,Leutwyler:1980tn}. They grow rapidly with
the principal quantum  number and result in a large theoretical uncertainty even
in the case of the $\Upsilon(1S)$ state. On the other hand for high moments  the
contribution of the sum to Eq.~(\ref{eq::momentnr}) significantly exceeds the
one of the integral and one may expect the nonperturbative effects to become as
important as for the bound state parameters. Indeed, for a large moment number
$n$ the nonperturbative correction to the sum rules result for the bottom quark
mass scales as $n^2\Lambda_{QCD}^4/m_b^4$, and for $n\sim 1/\alpha_s^2$  it is
parametrically as large as the nonperturbative  contribution to the first of
Eqs.~(\ref{eq::upsilonmasswidth}). However, the explicit calculations
\cite{Voloshin:1987rp,Voloshin:1995sf} show that the moments get  a negligible
contribution from the momentum region of order $\Lambda_{QCD}$  for $n\lessim
20$, though the nonperturbative contribution in  this case is suppressed
numerically rather than parametrically. Thus for  $n\sim 1/\alpha_s^2$ the
moments are significantly less sensitive to the long-distance phenomena than the
bound state parameters. This is in agreement with a general argument that the
moments are inclusive Euclidean quantities, which is valid even when the bound
state contributions to the spectral representation is large. The
evaluation of the perturbative corrections to the parameters $E_n$ and $Z_n$ is,
therefore, crucial for high-order analysis of the nonrelativistic sum  rules. In
the next section we present the complete ${\cal O}(\alpha_s^3)$ result for these
quantities. 
\subsection{Heavy quarkonium mass and leptonic width to
${\cal O}(\alpha_s^3)$}
\label{sec::quarkonium}
We parameterize  the perturbative series for the resonance mass and leptonic
width as follows
\begin{equation}
E_n=E_n^C\sum_{m=0}^\infty \left({\alpha_s\over \pi}\right)^me_n^{(m)},\qquad
Z_n=|\psi^C_n(0)|^2\sum_{m=0}^\infty \left({\alpha_s\over \pi}\right)^mz_n^{(m)}\,,
\label{eq::energyzseries}
\end{equation}
where $\alpha_s\equiv \alpha^{(n_l)}_s(\mu)$ is the $\overline{\rm MS}$
renormalized coupling with $n_l$ light-quark flavors and
$e_n^{(0)}=z_n^{(0)}=1$. The first two coefficients of the series for the
binding energy are well known \cite{Pineda:1997hz} and listed in the Appendix A.
The third-order term can be decomposed according to the powers of the logarithm
\begin{eqnarray}
e^{(3)}_n&=&\left({1331\over 2}-121n_l+{22\over 3}n_l^2-{4\over 27}n_l^3\right)L_n^3
+{\delta^{(2)}_e}(n)L_n^2
\nonumber\\
&+&\left({9\over 2}+{424\over 9n}-{32\over 9n^2}\right)\pi^2L+{\delta^{(1)}_e}(n)L_n
+{\delta^{(0)}_e}(n)\,,
\label{eq::e3}
\end{eqnarray}
where $L_n=\ln\left({n\mu/\alpha_sC_Fm_q}\right)$ and $L=\ln\left(\mu/m_q\right)$.
The  full analytical expression of the coefficients in Eq.~(\ref{eq::e3}) for
arbitrary $n$ \cite{Penin:2005eu,Beneke:2005hg} is too cumbersome  and  we only
present their values for the six lowest states.  For the ground state they read
\cite{Penin:2002zv}
\begin{eqnarray}
{\delta^{(2)}_e}(1)&=&{4521\over 2}-{10955\over 24}n_l
+{1027\over 36}n_l^2-{5\over 9}n_l^3\,,
\nonumber\\
{\delta^{(1)}_e}(1)&=&{247675\over 96}+{26897\over 108}\pi^2
+{3025\over 2}\zeta(3)-{99\over 16}\pi^4
+\left(-{166309\over 288}-{5095\over 162}\pi^2-{902\over 3}\zeta(3)
\right.
\nonumber\\
&+&\left.{3\over 8}\pi^4\right)n_l+\left({10351\over 288}
+{11\over 9}\pi^2+{158\over 9}\zeta(3)\right)n_l^2
+\left(-{50\over 81}-{2\over 81}\pi^2-{8\over 27}\zeta(3)\right)n_l^3\,,
\nonumber\\
{\delta^{(0)}_e}(1)&=&7362.11-1318.36\,n_l+75.2630\,n_l^2-1.25761\, n_l^3\,,
\label{eq::deltae1}
\end{eqnarray}
where $\zeta(z)$ is the Riemann zeta-function, $\zeta(3)=1.20206\ldots$. The
last coefficient incorporates the three-loop contribution to the  static
potential~\cite{Anzai:2009tm,Smirnov:2009fh}. For $n=2,\ldots,6$ the
coefficients are listed in the Appendix~B. The corrections to the leptonic width
up to ${\cal O}(\alpha_s^2)$ can be read off the
results~\cite{Melnikov:1998ug,Penin:1998kx,Czarnecki:1997vz,Beneke:1997jm} and
are given in the Appendix~A. The third-order term for general $n$ has only been
evaluated in the logarithmic approximation
\cite{Kniehl:1999mx,Kniehl:2002yv,Hoang:2003ns}. Recently the total third-order
correction for  $n=1$  has been published in a numerical form
\cite{Beneke:2014qea}. Below we  present the result for the excited states.  As it
follows from Eq.~(\ref{eq::energyzseries}), the ${\cal O}(\alpha_s^3)$
contribution consists of:
\begin{itemize}
\item[(i)] Interference of the leading and next-to-leading order corrections
from each of three factors in Eq.~(\ref{eq::zdef}).
\item[(ii)] The one-loop correction to the matching coefficient $d_v$
\cite{Luke:1997ys} and  recently completed three-loop corrections to matching
coefficient $c_v$ \cite{Marquard:2006qi,Marquard:2009bj,Marquard:2014pea}.
\item[(iii)] The corrections to the wave function
\cite{Penin:2005eu,Beneke:2005hg,Beneke:2007gj,Beneke:2013jia} due to the N$^3$LO
operators in the effective Hamiltonian
\cite{Anzai:2009tm,Smirnov:2009fh,Kniehl:2001ju,Kniehl:2002br} and due to the
multiple iterations of the next-to-leading  and NNLO operators.
\item[(iv)] The ultrasoft correction to the wave function \cite{Beneke:2007pj}
\end{itemize}
Combining all the above contribution we get the complete result for the
third-order term in the following form
\begin{eqnarray}
z^{(3)}_n&=&\left({6655\over 4}-{605\over 2}n_l+{55\over 3}n_l^2-{10\over 27}n_l^3\right)L_n^3
+\left(-{484\over 3}-{1406\over 27}\pi^2+\left({176\over 9}
+{140\over 81}\pi^2\right)n_l\right.
\nonumber \\
&-&\left.{16\over 27}n_l^2\right)L^2
+\left(-484-{346\over 9}\pi^2+\left({176\over 3}
+{140\over 27}\pi^2\right)n_l-{16\over 9}n_l^2\right)LL_n
\nonumber \\
&+&{\delta^{(2)}_z}(n)L_n^2+{{\delta'}^{(1)}_z}(n)L
+{\delta^{(1)}_z}(n)L_n+{\delta^{(0)}_z}(n)\,.
\label{eq::z3}
\end{eqnarray}
For the ground state we obtain the following values of the $n$-dependent
coefficients
\begin{eqnarray}
{\delta^{(2)}_z}(1)&=&{6809\over 2}-{50119\over 108}\pi^2+\left(-{37943\over 48}
+{15215\over 162}\pi^2\right)n_l+\left({3935\over 72}
-{55\over 9}\pi^2\right)n_l^2
\nonumber\\
&+&\left(-{7\over 6}+{10\over 81}\pi^2\right)n_l^3\,,
\nonumber\\
{{\delta'}^{(1)}_z}(1)&=&-{15553\over 27}-{26981\over 4860}\pi^2
-{2750\over 9}\zeta(3)+{770\over 81}\pi^4
-{2284\over 27}\pi^2\ln{2}
\nonumber\\
&+&\left({7468\over 81}+{337\over 54}\pi^2+{500\over 27}\zeta(3)
-{140\over 243}\pi^4+{56\over 27}\pi^2\ln{2}\right)n_l
+\left(-{260\over 81}+{16\over 81}\pi^2\right)n_l^2\,,
\nonumber\\
{\delta^{(1)}_z}(1)&=&{214891\over 144}-{7274327\over 38880}\pi^2
+{193985\over 48}\zeta(3)+{20735\over 10368}\pi^4
-{26\over 27}\pi^2\ln{2}
\nonumber\\
&+&\left(-{916993\over 1728}+{458711\over 3888}\pi^2-{30185\over 36}\zeta(3)
-{45445\over 15552}\pi^4+{28\over 9}\pi^2\ln{2}\right)n_l
+\left({76739\over 1728}\right.
\nonumber\\
&-&\left.{7045\over 648}\pi^2+{205\over 4}\zeta(3)+{55\over 216}\pi^4\right)n_l^2
+\left(-{80\over 81}+{227\over 972}\pi^2-{25\over 27}\zeta(3)
-{5\over 972}\pi^4\right)n_l^3\,,
\nonumber\\
{\delta^{(0)}_z}(1)&=&-3557(4)+310.22(2)\,n_l-2.83280\,n_l^2+0.0565322\, n_l^3\,,
\label{eq::deltaz1}
\end{eqnarray}
where the error in  $\delta^{(0)}_z$ is due to the numerical evaluation  of $c_v$
\cite{Marquard:2014pea}. The corresponding expressions for $n=2,\ldots,6$ are
listed in the Appendix~B.
\section{Numerical analysis}
\label{sec::numerics}
Now we are in a position to apply the method described in the previous section
for the determination of the bottom quark mass. We use the parameters of the six
$\Upsilon$ resonances listed in Table~\ref{tab::upsilonexp} as the experimental
input. The $5$th and $6$th resonances actually lie above the $B$-meson
production threshold and  their contribution does not represent the total
experimental spectral density in this region. We use this contribution only to
estimate the experimental uncertainty of our result. For the QED running
coupling we adopt the value   $\alpha(2m_b)=1.036\,\alpha$ \cite{alphaQED},
where $\alpha=1/137.036$ is the fine structure constant.

\begin{table}[t]
  \begin{center}
    \begin{tabular}{|c|c|c|c|}
 \hline
 $n$& 1 & 2 & 3\\
 \hline
 $M_{\Upsilon(nS)}$~(GeV)& 9.46030(26) & 10.02326(31) &  10.3552(5) \\
 \hline
 $\Gamma_{\Upsilon(nS)\to e^+e^-}$~(keV)& 1.340(18) & 0.612(11) & 0.443(8) \\
  \hline  \hline
 $n$& 4& 5& 6\\
 \hline
 $M_{\Upsilon(nS)}$~(GeV)& 10.5794(12) & 10.876(11)&  11.019(8)  \\
 \hline
 $\Gamma_{\Upsilon(nS)\to e^+e^-}$~(keV)& 0.272(29)  & 0.31(7)&  0.130(30)\\
  \hline
    \end{tabular}
    \caption{\label{tab::upsilonexp}
    Experimental values of the $\Upsilon$-resonance masses and leptonic widths
    \cite{Beringer:1900zz}.}
  \end{center}
\end{table}

On the theory side we use the complete ${\cal O}(\alpha_s^3)$ expression for the
contribution of the perturbative heavy-quarkonium  bound states below the $b\bar
b$ threshold given in  Section~\ref{sec::quarkonium} for $n_l=4$. The continuum
contribution from the above-threshold region  is strongly suppressed for high
moments and high accuracy of the theoretical approximation there is not
mandatory. Thus, without introducing significant error we emulate the N$^3$LO
spectral density for  $s>4m_b^2$  by rescaling the NNLO result \cite{Penin:1998zh}
\begin{equation}
R(s)={Z_1^{N^3LO}\over Z_1^{NNLO}}R^{NNLO}(s)\,.
\label{eq::rn3lo}
\end{equation}
To estimate the error introduced by this approximation we multiply the total
continuum contribution to the theoretical moments by  $1/2<\rho<2$, which
corresponds to the variation of its absolute value by factor four.

\begin{figure}[t]
  \begin{center}
    \begin{tabular}{c}
    \includegraphics[width=0.6\textwidth]{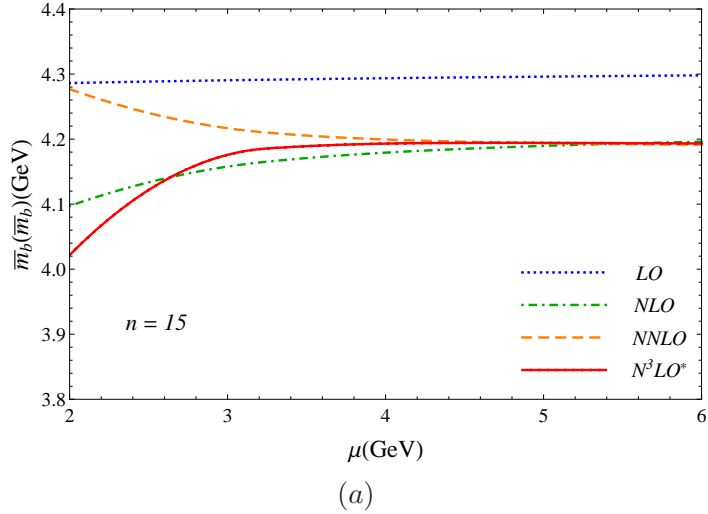}
      \\
      $(a)$ \\ \\
    \includegraphics[width=0.6\textwidth]{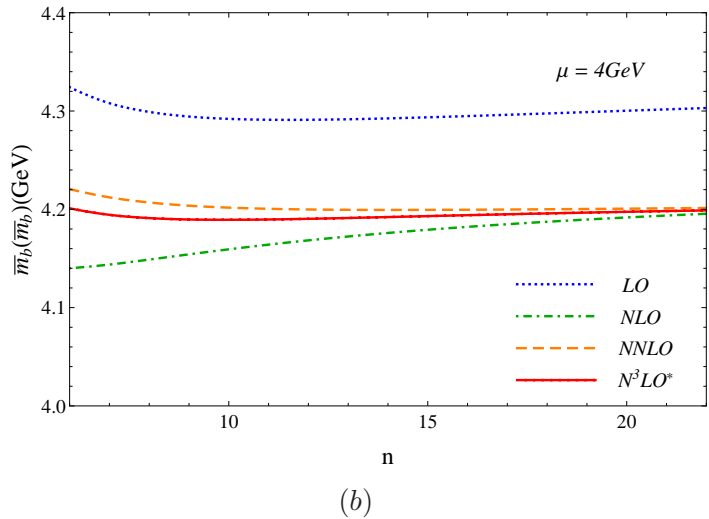}
      \\ $(b)$
    \end{tabular}
  \end{center}
  \caption{\label{fig::msbarmass} The bottom quark mass plotted $(a)$ as
  function of the renormalization scale for $n=15$ and $(b)$ as function of the
  moment number for $\mu = 4$~GeV in different orders of perturbation theory.
  The star in N$^3$LO$^*$ refers to the
  approximate character of Eqs.~(\ref{eq::rn3lo},~\ref{eq::r4renormalon}).}
\end{figure}

The sequence of the values of  the bottom quark pole mass $m_b$ extracted from
the sum rules order by order in perturbation theory does not converge well. This
is expected since the pole mass is widely believed not to be a good parameter of
 perturbative QCD due to infrared renormalon contribution (see
\cite{Beneke:1998ui} for a review and \cite{Bauer:2011ws} for a recent
high-order analysis). The perturbative behavior of the ``short-distance'' mass
parameter $\overline{m}_b(\mu)$ defined in  $\overline{\rm MS}$ renormalization
scheme is supposed to be much better.  Therefore we  convert the extracted pole
mass value into $\overline{m}_b(\overline{m}_b)$ according to the relation
\begin{equation}
m_b= \overline{m}_b(\overline{m}_b)\sum_nr^{(n)}
\left({\alpha_s(\overline{m}_b)\over \pi}\right)^n\,.
\label{eq::massconversion}
\end{equation}
where the coefficients $r^{(n)}$ have been evaluated  up to $n=3$
\cite{Chetyrkin:1999qi,Melnikov:2000qh}. To achieve the cancellation of
factorially growing terms associated with the infrared renormalon one has to
correlate perturbative approximations for the mass relation and the sum rules in
such a way that  the series~(\ref{eq::massconversion}) is truncated at one order
higher than the series~(\ref{eq::energyzseries}), {\it i.e.} the one-loop mass
relation is used with the Coulomb approximation for the moments, and so on
\cite{Hoang:1998ng,Kiyo:2000fr}. Our analysis therefore requires the four-loop
coefficient $r^{(4)}$. Since the exact value of this coefficient is not yet
available we use the renormalon-based estimate \cite{Pineda:2001zq}
\begin{equation}
r^{(4)}_{ren}\approx 1346\,.
 \label{eq::r4renormalon}
\end{equation}
This method  reproduces the value of the three-loop coefficient  $r^{(3)}$ with
the precision better than  one part in a thousand. We take  the
difference between Eq.~(\ref{eq::r4renormalon}) and the large-$\beta_0$
prediction $r^{(4)}_{\beta_0}\approx 1325$  \cite{Beneke:1998ui} as a
conservative estimate of its uncertainty.

An important issue of the numerical analysis of high moments is whether the
factor  $1/(2m_b+E_m)^{2n+1}$ in Eq.~(\ref{eq::momentnr}) should be expanded
about  $1/(2m_b+E^C_m)^{2n+1}$. Formally all the terms of this expansion beyond
the N$^3$LO are suppressed according to the  $\sqrt{n}\sim 1/\alpha_s$ power
counting and can be neglected in our approximation. Such an expansion, however,
merely violates the cancellation of the large perturbative corrections related to
the infrared renormalon in the series for the binding
energy~(\ref{eq::energyzseries}) and for the pole mass~(\ref{eq::massconversion})
\cite{Melnikov:1998ug,Hoang:1999ye,Beneke:1999fe,Hoang:1998ng,Kiyo:2000fr,Pineda:2001zq}.
This  spoils the convergence of the resulting series for $\overline{m}_b$  since
the factorial growth of the coefficients $e^{(k)}_m$ beats the $1/\sqrt{n}$
suppression. To ensure the cancellation to all orders in $1/\sqrt{n}$ we first
extract the value of the pole mass keeping the above factor unexpanded and then
convert it into $\overline{m}_b$ to the required order of perturbation theory.

The result for the $\overline{\rm MS}$  bottom quark mass is shown in
Figs.~\ref{fig::msbarmass}(a) and ~\ref{fig::msbarmass}(b) in the different
orders of perturbation theory as functions of the moment number and the
renormalization scale.  For the numerical estimates we use the moments in the
interval $10\grtsim n \grtsim 20$,  where the nonrelativistic perturbative
approximation for the spectral density is valid, the nonperturbative effects
are under control, and the result is almost insensitive to the continuum
contribution to the theoretical moments. The renormalization
scale is varied  in a physically motivated interval between the soft scale
$\mu\sim\alpha_sm_b$ and the hard scales $\mu\sim m_b$. We take
$\alpha_s(M_z)=0.1184\pm 0.0007$ \cite{Beringer:1900zz} as an input and run it
down to $\mu=m_b$ with the four-loop beta-function \cite{Chetyrkin:2000yt}. To
convert $\alpha_s(m_b)$ into $\alpha_s(\mu)$ used in our numerical analysis we
correlate the order of the renormalization group evolution of the strong
coupling constant with the perturbative expansion for the sum rules so that the
one-loop running is used in the leading approximation, an so on. To ensure  the
renormalon cancellation we reexpress Eq.~(\ref{eq::massconversion}) through
$\alpha_s(\mu)$ and use the same renormalization scale both for the sum rules
and the mass relation.

\begin{table}[t]
  \begin{center}
    \begin{tabular}{|c|c|c|c|c|c|c|c|}
     \hline
     $\Delta_{exp}$& $\Delta_{\alpha_s}$ & $\Delta_\rho $ &
     $\Delta_{r^{(4)}} $& $\Delta_{n}$ & $\Delta_{p.t.}$& $\Delta_{n.p.}$&
     $\Delta_{m_c} $ \\
     \hline
     $2.3$& $1.9$ & $4.2$ & $2.2$& $3.4$ & $2.1$ & $0.8$ & $5.0$\\
     \hline
    \end{tabular}
    \caption{\label{tab::errorbars}
     Different contributions to the uncertainty of
     $\overline{m}_b(\overline{m}_b)$ in MeV.}
  \end{center}
\end{table}

\begin{figure}[t]
  \begin{center}
    \begin{tabular}{c}
   \includegraphics[width=0.63\textwidth]{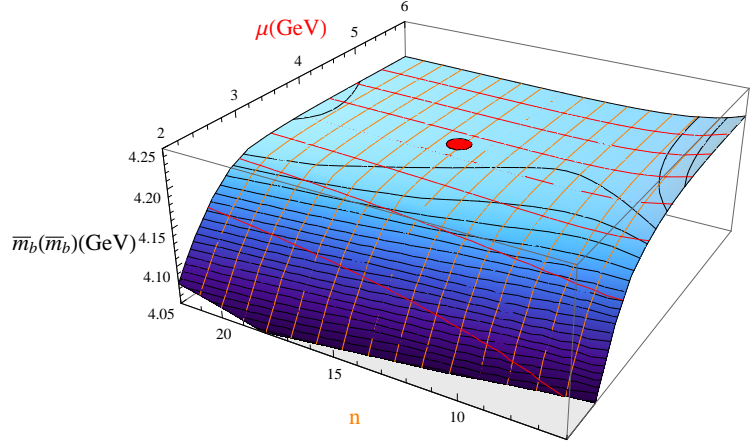}
      \\
      $(a)$ \\ \\ \\
    \includegraphics[width=0.63\textwidth]{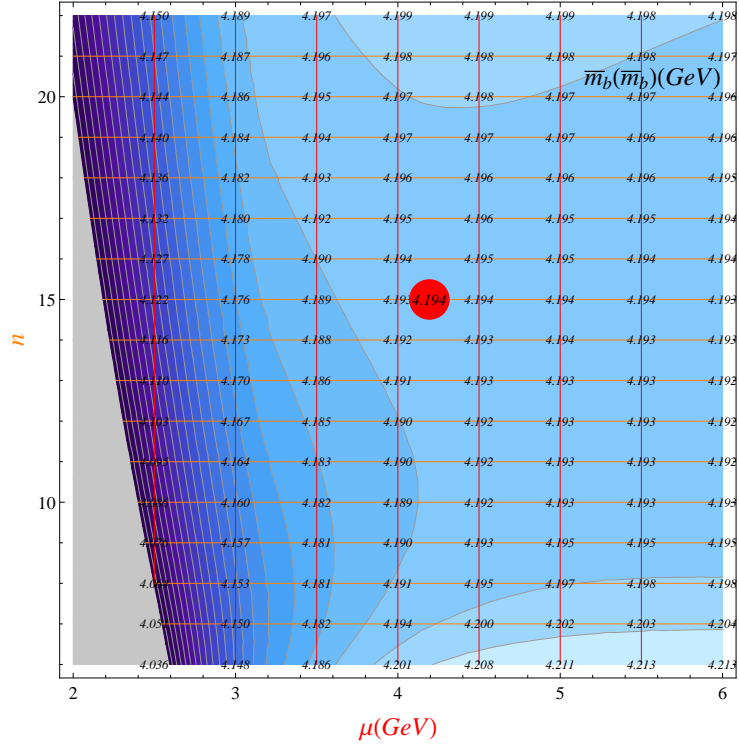}
      \\ $(b)$
    \end{tabular}
  \end{center}
  \caption{\label{fig::msbarmass2dim} Three-dimensional $(a)$ and contour $(b)$
  plots of the  bottom quark mass as function of the renormalization scale and
  moment number in the N$^3$LO approximation. The red blob corresponds to the
  central value of our estimate.}
\end{figure}

\begin{figure}[t]
  \begin{center}
    \begin{tabular}{c}
   \includegraphics[width=0.6\textwidth]{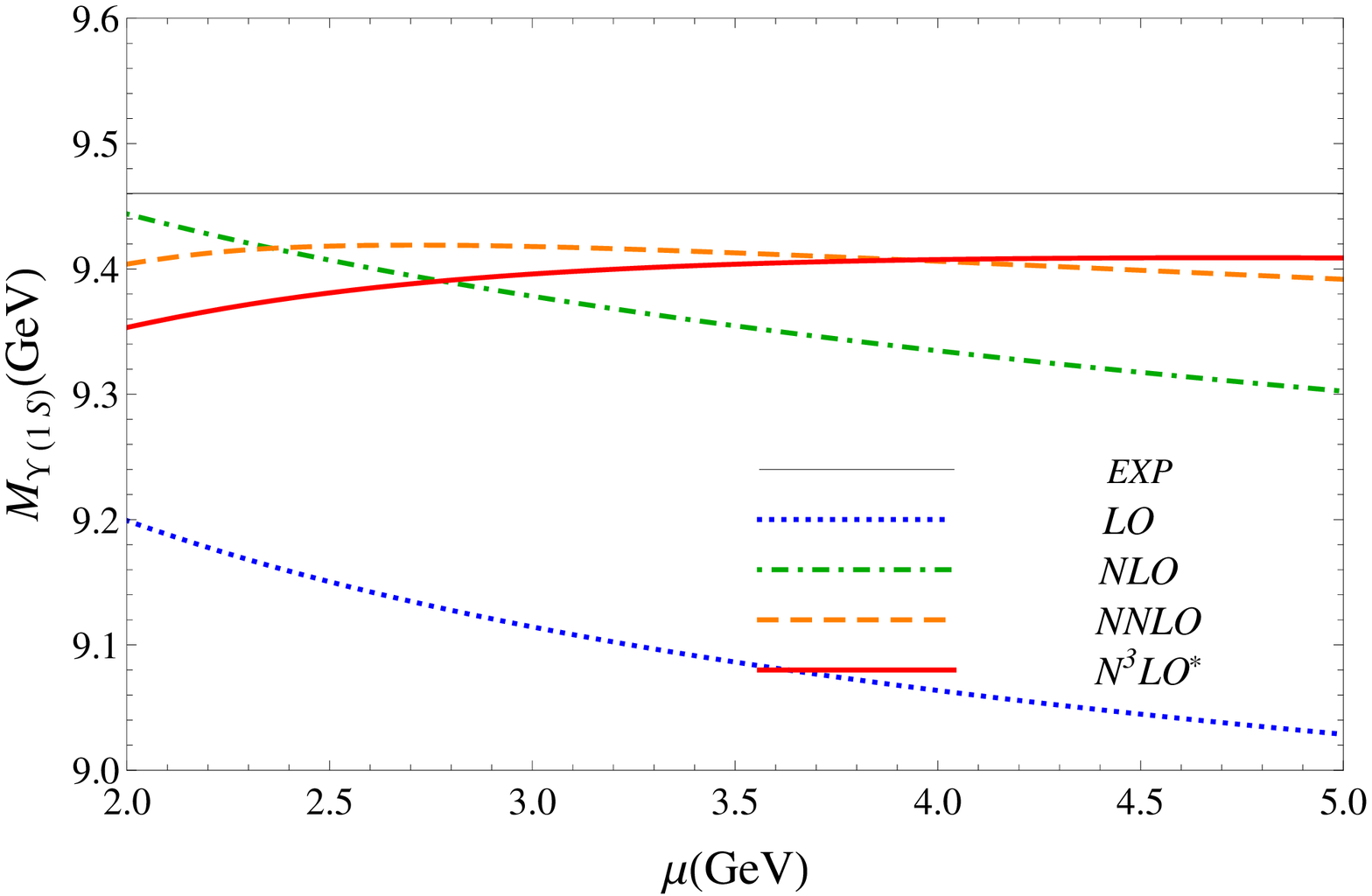}
      \\
      $(a)$ \\ \\
   \includegraphics[width=0.6\textwidth]{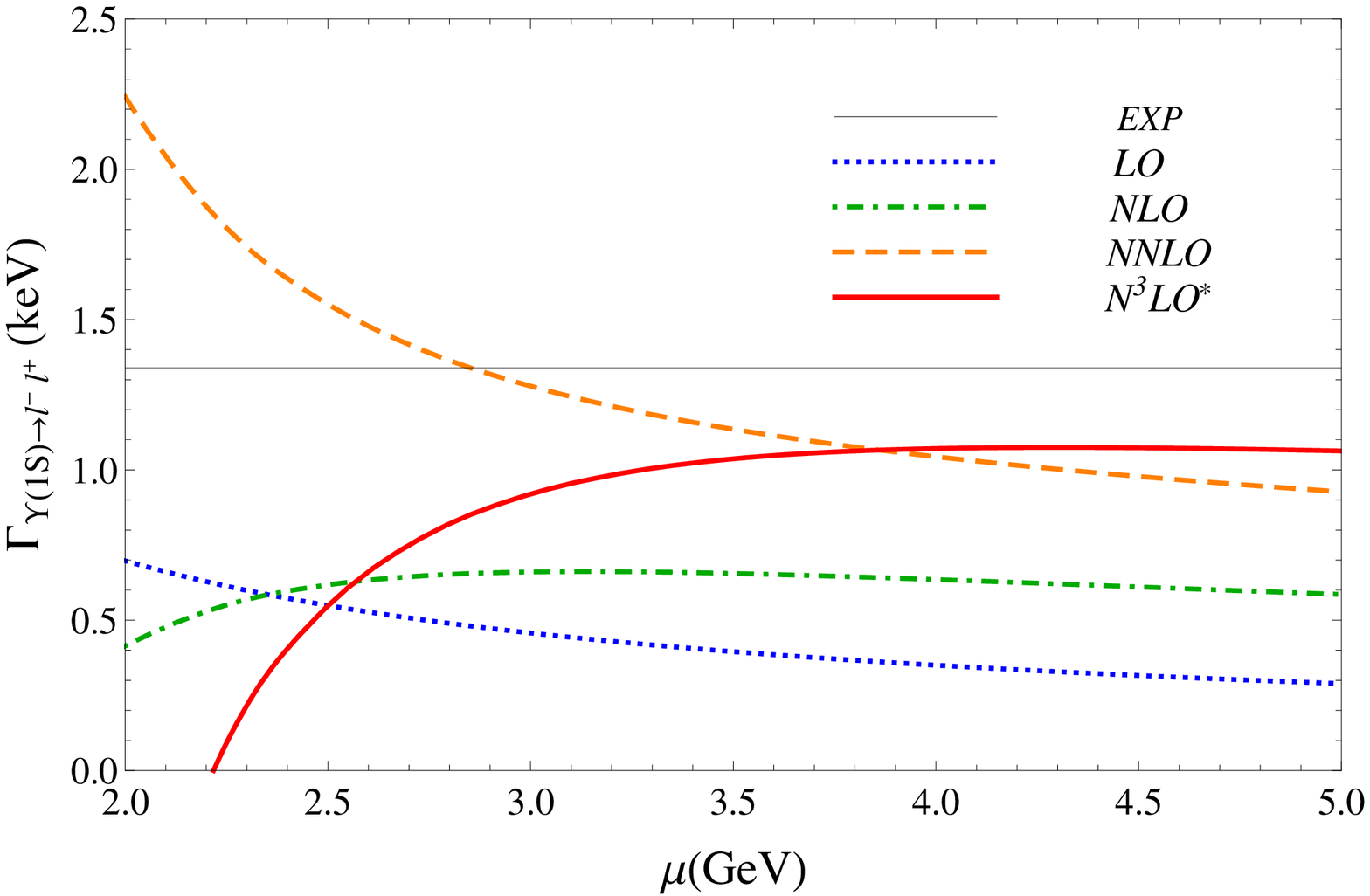}
      \\ $(b)$
    \end{tabular}
  \end{center}
  \caption{\label{fig::upsilonmasswidth} The   $\Upsilon(1S)$  mass $(a)$ and
  leptonic width   $(b)$ plotted as functions of the renormalization scale
  in different orders of perturbation theory for $\alpha_s(M_Z)=0.1184$ and
  and $\overline{m}_b(\overline{m}_b)=4.194$~GeV.
  }
\end{figure}

A stable perturbative result is achieved  for $n\grtsim 10$ and $ \mu\grtsim
3.5$~GeV, see Fig.~\ref{fig::msbarmass2dim}. We take
$\overline{m}_b(\overline{m}_b)=4.194$ as a central value of our estimate. It
corresponds to  $n=15$ in the center of the allowed interval and to
the renormalization scale $\mu=\overline{m}_b$, which belongs to the stability
plateau and provides  $n\alpha_s^2\approx 0.8$ in agreement with the power
counting rules. The uncertainty  budget of our estimate is summarized in
Table~\ref{tab::errorbars}. The experimental part of the uncertainty
$\Delta_{exp}$ accounts for both the  error bars in the measured values of the
resonance mass and width, and the contribution from the region above the
$B$-meson production threshold. We estimate the latter by the size of the $5$th
and $6$th $\Upsilon$-resonance contribution to the experimental moments. The
uncertainty $\Delta_{\alpha_s}$ corresponds to the error in the input value of
$\alpha_s(M_Z)$. The quantities $\Delta_{\rho}$ and $\Delta_{r^{(4)}}$ account
for the approximate character of the
Eqs.~(\ref{eq::rn3lo},~\ref{eq::r4renormalon}). The value of $\Delta_{\rho}$
is given by a half  of the variation of the result with the parameter $\rho$
changing from $1/2$ to $2$. In the same way $\Delta_{n}$ is given by
a half of the variation of the result with the moment number spanning the
interval $10< n < 20$.  We take one half of the third-order correction as an
estimate of the uncertainty $\Delta_{p.t.}$ introduced by  truncation  of the
perturbative  expansion. This procedure can be verified through the N$^3$LO.
Indeed, the numerical series for the bottom quark mass reads
\begin{equation}
\overline{m}_b(\overline{m}_b)=4.294\left(1_{LO}-0.0262_{NLO}+0.0038_{NNLO}
-0.0010_{N^3LO}+\ldots\right)\,,
 \label{eq::convergence}
\end{equation}
where the third-order correction amounts only about a quarter of the
second-order term. Each individual contribution to the total uncertainty is
computed with all other parameters fixed at  their central values and  at the
normalization scale $\mu=\overline{m}_b$. We refrain from using the scale
dependence for the uncertainty estimate since this procedure strongly depends on
the low boundary of the allowed scale variation which cannot be unambiguously
defined. We therefore restrict the analysis only to the large stability region
where the variation of the result due to the change of the scale is much smaller
than our estimate of perturbative uncertainty given above. The choice of the
hard renormalization scale may look ambiguous since for large $n$ the soft scale
$m_b/\sqrt{n}$ and the ultrasoft scale $m_b/n$ are involved. As a consequence the
coefficients of the series get contributions enhanced by logarithm  of a
scale ratio proportional to $\ln{n}$, which may affect the convergence of the
perturbative expansion.  The  logarithmic terms cannot be completely eliminated
by adjusting the renormalization scale of $\alpha_s$ but can be resummed to
all orders through the effective theory renormalization group
\cite{Pineda:2006gx,Hoang:2012us,Penin:2004ay}. However, for our choice of the
moments $\ln{n}<3$, {\it i.e.} the asymptotic regime is not yet reached. In fact
for such $n$ the logarithmic terms do not saturate the coefficients of the
series numerically and do not have to be distinguished from the nonlogarithmic
contributions. This justifies our choice of the renormalization scale dictated
solely by the convergence of the perturbation theory.

The moments get also a contribution from the  nonperturbative scale
$\Lambda_{QCD}$.  Within the  operator product expansion it is given by a series
in $n\Lambda^2_{QCD}/m_b^2$. The leading nonperturbative contribution to the
high moments due to the gluon condensate turns out to be numerically suppressed
\cite{Voloshin:1995sf}. For $n=15$ and $\langle{\alpha_s\over\pi} G^2 \rangle
\approx 0.012~{\rm GeV}^4$ the corresponding  correction to the bottom quark
mass is about $-0.8$~MeV. We take this value as the nonperturbative uncertainty
$\Delta_{n.p.}$ of our result. Though the nonperturbative uncertainty  rapidly
increases with the moment number, the perturbative one is suppressed
for higher moments and their sum does not significantly change over the whole
range of $n$ considered in the paper. One may be concerned  that for large $n$ the
ultrasoft scale $m_b/n$ approaches $\Lambda_{QCD}$, which questions the
perturbative treatment of the ultrasoft contribution.  However, the result of
Ref.~\cite{Voloshin:1995sf} ensures that even for $n=20$ the moments do not get
a sizable contribution from the gluonic field fluctuations at the
scale $\Lambda_{QCD}$ and the perturbative description of the moments is
justified.

\subsection{Charm mass effect}
So far we considered the charm quark to be massless. Since its mass $m_c$
is not much smaller  than the soft scale, the charm quark mass
effect on the effective potential and the bound state dynamics may not be
negligible \cite{Eiras:2000rh,Melles:2000dq}. This effect has been analyzed
in the context of $\Upsilon$ sum rules through  the NNLO \cite{Hoang:2000fm}.
By using the results  of  Ref.~\cite{Hoang:2000fm} for the charm quark mass
correction to the moments and to the mass relation,
for $\overline{m}_c(\overline{m}_c) \approx 1.3$~GeV \cite{Chetyrkin:2010ic}
and a set of input parameters similar to the one adopted in this section we get
a negative shift of approximately $25$~MeV in the extracted value of
$\overline{m}_b(\overline{m}_b)$ with an estimated error $\Delta_{m_c}=5$~MeV.
We incorporate this correction in the analysis. Our final prediction for the
bottom quark mass reads
\begin{equation}
\overline{m}_b(\overline{m}_b)=4.169\pm 0.008_{th}\pm 0.002_{\alpha_s}\pm 0.002_{exp} \,,
\label{eq::final}
\end{equation}
where the theoretical error corresponds to $\Delta_\rho$,  $\Delta_{r^{(4)}}$,
$\Delta_{n}$, $\Delta_{p.t.}$, $\Delta_{n.p.}$,  and  $\Delta_{m_c}$ added up
in quadrature.

\subsection{$\Upsilon(1S)$  mass and leptonic width}
\label{sec::masswidth}
Though significant nonperturbative effects are expected in the QCD analysis of
the $\Upsilon$-resonance mass and width, it is instructive to figure out  how
perturbative QCD results \cite{Penin:2002zv,Beneke:2014qea} reproduce the
experimental data for the lowest $\Upsilon(1S)$ state, where the nonperturbative
contribution is minimal. From the result of the previous section we get the
following numerical series for the ground state mass and width for $m_c=0$
\begin{eqnarray}
M_{\Upsilon(1S)}^{p.t.}&=&2m_b+E_1^C\left[1+\left(2.653\,L_1+3.590\right)\alpha_s
+\left(5.277\,L_1^2+12.066\,L_1+19.524\right)\alpha_s^2\right.
\nonumber \\
&+&\left.\left(9.332\,L_1^3+27.593\,L_1^2+15.297\,L+78.375\,L_1+103.605\right)\alpha_s^3
+\ldots\right]\,,
\label{eq::mass1s}
\end{eqnarray}
\begin{eqnarray}
\Gamma_{\Upsilon(1S)\to l^+l^-}^{p.t.}&=&\Gamma_{\Upsilon(nS)\to l^+l^-}^{LO}
\left[1+\left(3.979\,L_1-2.003\right)\alpha_s+\left(10.554\,L_1^2-7.437\,L\right.\right.
\nonumber \\
&-&\left.6.514\,L_1+11.188\right)\alpha_s^2
+\left(23.330\,L_1^3-17.361\,L^2-14.594\,L_1L-23.125\,L_1^2\right.
\nonumber \\
&-&\left.\left.4.339\,L+80.603\,L_1-76.033\right)\alpha_s^3
+\ldots\right]\,,
\label{eq::width1s}
\end{eqnarray}
in agreement with \cite{Penin:2002zv,Beneke:2014qea}. In
Fig.~\ref{fig::upsilonmasswidth} we plot $M_{\Upsilon(1S)}$ and
$\Gamma_{\Upsilon(1S)\to e^+e^-}$ evaluated according to
Eqs.~(\ref{eq::mass1s},~\ref{eq::width1s}) as functions of the renormalization
scale with a fixed value of  $\overline{m}_b(\overline{m}_b)=4.194$~GeV as an
input parameter. The plots clearly indicate stabilization of the perturbative
expansion at $\mu\sim \overline{m}_b$. Evidently the inclusion of the high-order
corrections  improves the accuracy of  the perturbative QCD prediction. The
convergence of the sum rules series~(\ref{eq::convergence}), however, is by far
superior to the series~(\ref{eq::mass1s}).

The ${\cal O}(\alpha_s^3)$ approximation is in a rather good agreement with the
experimental data for the resonance mass (width). The difference of about
$60$~MeV ($0.3$~keV)  quantitatively agrees  with an estimate of the
nonperturbative contribution of the gluon condensate \cite{Pineda:1996uk}. The
inclusion of the charm mass effects increases the difference between the
perturbative result and the measured $\Upsilon(1S)$ mass to approximately
$90$~MeV since the reduction of the binding energy~\cite{Hoang:2000fm} does not
fully compensate the negative corrections to the input  value of
$\overline{m}_b$~(\ref{eq::final}).  The  interpretation of the nonperturbative
contribution to Eqs.~(\ref{eq::mass1s},~\ref{eq::width1s}) in this case is more
ambiguous \cite{Beneke:2014qea} but still consistent with the gluon condensate.
This validates our estimate of the nonperturbative correction to the sum rules,
which is also based on the gluon condensate contribution. As it has been pointed
out in Section~\ref{sec::sumrules},  for $n\sim 1/\alpha_s^2$  the
nonperturbative correction to Eq.~(\ref{eq::final}) is not parametrically
suppressed in comparison to the first equation of~(\ref{eq::upsilonmasswidth}).
Nevertheless, numerically the gluon condensate contribution to the sum rules
result for $\overline{m}_b$ is almost two orders of magnitude smaller than the
nonperturbative correction to the resonance mass, clearly showing the 
short-distance nature of the moments.

\section{Summary and discussion}
\label{sec::summary}

\begin{table}[t]
  \begin{center}
    \begin{tabular}{|c|c|c|c|}
     \hline
      Reference & Method & Approximation & $\overline{m}_b(\overline{m}_b)$~(GeV) \\  \hline
      Ref.~\cite{Penin:2002zv} & $\Upsilon(1S)$ mass & ${\cal O}(\alpha_s^3)$  & $4.346\pm 0.070$  \\ \hline
      Ref.~\cite{Beneke:2005hg} & $\Upsilon(1S)$ mass & ${\cal O}(\alpha_s^3)$  & $4.25\pm 0.08$  \\ \hline
      Ref.~\cite{Pineda:2006gx}   & high moments & partial NNLL & $4.190\pm 0.060$ \\ \hline
      Ref.~\cite{Hoang:2012us}   & high moments & partial NNLL   &  $4.235\pm 0.055$ \\ \hline
      Ref.~\cite{Chetyrkin:2010ic}  & low moments & ${\cal O}(\alpha_s^3)$  & $4.163\pm 0.016$ \\ \hline
      This work   & high moments & ${\cal O}(\alpha_s^3)$ & $4.169\pm 0.009$\\ \hline
    \end{tabular}
    \caption{\label{tab::comparison}
     The results of the bottom quark mass determination from $\Upsilon$ family
     properties beyond the NNLO. In the last line all the errors given in
     Table.~\ref{tab::errorbars} are added up in quadrature.}
  \end{center}
\end{table}

The main result of this work is the new value of the bottom quark
mass~(\ref{eq::final}) from the ${\cal O}(\alpha_s^3)$ analysis of the
nonrelativistic  $\Upsilon$-sum rules. In Table~\ref{tab::comparison} we
confront Eq.~(\ref{eq::final}) with the existing results of the bottom quark
mass determination from $\Upsilon$ phenomenology  beyond the NNLO of
perturbation theory. In Refs.~\cite{Penin:2002zv,Beneke:2005hg} the bottom quark
mass has been obtained from the ${\cal O}(\alpha_s^3)$ approximation for
$M_{\Upsilon(1S)}$. A relatively large value of the bottom quark mass reported
in Ref.~\cite{Penin:2002zv} is due to the choice of the ``physical'' soft
renormalization scale $\mu=2.7$~GeV  natural for the bound state dynamics.
However, the perturbative expansion becomes unstable at such a low scale ({\it
cf.} Fig.~\ref{fig::upsilonmasswidth}(a)). For $\mu=4.20$~GeV  the
analysis~\cite{Penin:2002zv} gives the value
$\overline{m}_b(\overline{m}_b)=4.22\pm 0.07$ consistent with
Refs.~\cite{Beneke:1999fe,Beneke:2005hg} and Eq.~(\ref{eq::final}).

The high-moment sum rules have been considered in a context of the effective
theory renormalization group in Refs.~\cite{Pineda:2006gx,Hoang:2012us}. Both
analyses involve partial resummation of the next-to-next-to-leading logarithms
(NNLL) of the form $\alpha_s^{m+2}\ln^{m}\alpha_s$ for all $m$.\footnote{The
complete NNLL result is available only for the spin dependent part of the
quarkonium production and annihilation rates \cite{Penin:2004ay}.} The result of
Ref.~\cite{Pineda:2006gx} agrees with Eq.~(\ref{eq::final}) within the error
bars. Though the renormalization group resummation improves the behavior of the
perturbative expansion especially at a low renormalization scale, the
logarithmic terms do not dominate the perturbative
series~(\ref{eq::energyzseries}) and cannot be used for a precise quantitative
estimate of the third-order corrections. As a consequence, the theoretical error
of the NNLL approximation is significantly larger than the one of the N$^3$LO
result.

The most accurate value of the bottom  quark mass up to date has been reported
in Ref.~\cite{Chetyrkin:2010ic} and is obtained from the relativistic sum rules
at ${\cal O}(\alpha_s^3)$. The central value  given in
Table~\ref{tab::comparison} corresponds to  $n=2$. The error estimate includes
$\pm 10$~MeV due to uncertainty of the experimentally measured cross section,
$\pm 12$~MeV due to the  input value  $\alpha_s(M_Z)=0.1189\pm0.002$, and the
theoretical uncertainty $\pm 3$ estimated by the variation of the
renormalization scale. The overall accuracy of the relativistic sum rules is
comparable to Eq.~(\ref{eq::final}) for a given interval of  $\alpha_s(M_Z)$.
However for the low moments the experimental error clearly dominates the
theoretical one. Our result is in a  perfect agreement with the analysis
\cite{Chetyrkin:2010ic}. The amazing agreement of two approaches based on
significantly different theoretical and experimental input boosts our confidence
in the result and, in particular, in the uncertainty assessment.
\acknowledgments
We thank to M.~Steinhauser for carefully reading the manuscript and many useful
comments. We are grateful to M.~Beneke, A.~Maier, J.~Piclum and T.~Rauh for the
cross-checks and discussion of the results. This work was supported  by NSERC
and the Alberta Ingenuity Foundation. The work of AP is supported in part by
Mercator DFG grant.
\newpage
\section{Appendix~A}
The first and the second-order coefficients of the
series~(\ref{eq::energyzseries}) for general $n$ read
\cite{Pineda:1997hz,Melnikov:1998ug,Penin:1998kx,Beneke:2005hg}
\begin{eqnarray}
e^{(1)}_n&=&\beta_0\left(L_n+S_1(n)\right)+{31\over 18}C_A-{10\over 9}n_lT_F\,,
\label{eq::e1}
\\[5mm]
e^{(2)}_n&=&{3\over 4}\beta_0^2L_n^2+\left[{\beta_1\over 4}
+\left({3\over 2}S_1(n)-{1\over 2}\right)\beta_0^2
+\left({31\over 12}C_A-{5\over 3}n_lT_F\right)\beta_0\right]L_n
+{S_1(n)\over 4}\beta_1
\nonumber \\
&+&\left[{\pi^2\over 24}+{\zeta(3)\over 2}n-\left({1\over 2}+{1\over 2n}\right)S_1(n)
+{3\over 4}S^2_1(n)+S_2(n)-{n\over 2}S_3(n)\right]\beta_0^2
\nonumber \\
&+&\left[\left({31\over 12}S_1(n)-{31\over 36}\right)C_A
+\left({5\over 9}-{5\over 3}S_1(n)\right)n_lT_F\right]\beta_0
\nonumber \\
&+&\left({221\over 54}+{\pi^2\over 2}-{\pi^4\over 32}+{11\over 12}\zeta(3)\right)C_A^2
-\left({403\over 108}+{7\over 3}\zeta(3)\right)n_lT_FC_A+{\pi^2\over n}C_FC_A
\nonumber \\
&+&{25\over 27}n_l^2T_F^2+\left(2\zeta(3)-{55\over 24}\right)n_lT_FC_F
+\left({2\over 3}-{11\over 16 n}\right){\pi^2\over n}C_F^2\,,
\label{eq::e2}
\\[5mm]
z^{(1)}_n&=&{3\over 2}\beta_0L_n+\left[{S_1(n)\over 2}
-{1\over 2}+\left(S_2(n)-{\pi^2\over 6}\right)n\right]\beta_0
+{31\over 12}C_A-{5\over 3}n_lT_F-4C_F\,,
\label{eq::z1}
\\[5mm]
z^{(2)}_n&=& {3\over 2}\beta_0^2L_n^2
+\left[{3\over 8}\beta_1+\left(S_1(n)-{7\over 4}+\left(2S_2(n)
-{\pi^2\over 3}\right)n\right)\beta_0^2+\left({31\over 6}C_A
-{10\over 3}n_lT_F\right.\right.
\nonumber \\
&-&6C_F\bigg)\beta_0+\left.\pi^2C_FC_A+{2\over 3}\pi^2C_F^2\right]L_n
-\left[2C_F\beta_0+\pi^2C_FC_A+{2\over 3}\pi^2C_F^2\right]L
\nonumber \\
&+&\left[{S_1(n)\over 8}-{1\over 8}+\left({S_2(n)\over 4}
-{\pi^2\over 24}\right)n\right]\beta_1+\left[{1\over 4}
+{\pi^2\over 16}+\left({\pi^2\over 12}+{5\over 4}\zeta(3)\right)n
+{\pi^4\over 144}n^2\right.
\nonumber \\
&+&\left({S_1(n)\over 2}+nS_2(n)-{5\over 4}-{3\over 4n}
-{\pi^2\over 6}n\right)S_1(n)+\left({1\over2}-{n\over 2}
-{\pi^2\over 12}n^2+{n^2\over 4}S_2(n)\right)S_2(n)
\nonumber \\
&+&\left.{7\over 4}nS_3(n)-{5\over 4}n^2S_4(n)-{3\over 2}nS_{2,1}(n)
+n^2S_{3,1}(n)\right]\beta_0^2+\left[\left({31\over 18}S_1(n)\right.\right.
\nonumber \\
&+&\left.\left.{31\over 9}nS_2(n)-{217\over 72}-{31\over 54}\pi^2n\right)C_A
+\left({35\over 18}+{10\over 27}\pi^2n-{10\over 9}S_1(n)
-{20\over 9}nS_2(n)\right)n_lT_F\right.
\nonumber \\
&+&\left.\left(2+{2\over 3}\pi^2n-2S_1(n)-4nS_2(n)\right)C_F\right]\beta_0
+\left[{6265\over 864}+{3\over 4}\pi^2-{3\over 64}\pi^4
+{11\over 8}\zeta(3)\right]C_A^2
\nonumber \\
&-&\left[{1519\over 216}+{7\over 2}\zeta(3)\right]n_lT_FC_A
+\left[\left({179\over 72}-{5\over 3}\ln{2}+{2\over n}\right)\pi^2
-{523\over 36}-{13\over 2}\zeta(3)\right.
\nonumber \\
&-&\pi^2S_1(n)\bigg]C_FC_A+{50\over 27}n_l^2T_F^2+
\left({641\over 144}+3\zeta(3)\right)n_lT_FC_F
+\left({44\over 9}-{4\over 9}\pi^2\right)T_FC_F
\nonumber \\
&+&\left[{39\over 4}+\left(2\ln{2}-{35\over 18}+{4\over 3n}
-{37\over 24n^2}\right)\pi^2-\zeta(3)-{2\over 3}\pi^2S_1(n)\right]C_F^2\,,
\label{eq::z2}
\end{eqnarray}
where $\beta_i$  are the coefficients of the beta-function,
$\beta_0={11\over 3}C_A-{4\over 3}n_lT_F$, {\it etc.}, $C_A=N_c$,  $T_F={1/2}$,
$S_i(n)=\sum_{m=1}^n{1/m^i}$ are the harmonic sums and $S_{i,j}(n)=\sum_{m=1}^n{S_j(m)/m^i}$
are the nested harmonic sums.

\section{Appendix~B}
The third-order coefficients of the series for the $n=2,\ldots,6$ binding energy read \cite{Penin:2005eu,Beneke:2005hg}
\begin{eqnarray}
{\delta^{(2)}_e}(2)&=&{13035\over 4}-{15311\over 24}n_l
+{1423\over 36}n_l^2-{7\over 9}n_l^3\,,
\nonumber\\
{\delta^{(1)}_e}(2)&=&{448711\over 96}+{25171\over 108}\pi^2
+{5687\over 2}\zeta(3)-{99\over 16}\pi^4
+\left(-{289057\over 288}-{4733\over 162}\pi^2-{1628\over 3}\zeta(3)
\right.
\nonumber\\
&+&\left.{3\over 8}\pi^4\right)n_l+\left({6013\over 96}
+{11\over 9}\pi^2+{290\over 9}\zeta(3)\right)n_l^2
+\left(-{92\over 81}-{2\over 81}\pi^2-{16\over 27}\zeta(3)\right)n_l^3\,,
\nonumber\\
{\delta^{(0)}_e}(2)&=&12043.4-2283.40\,n_l+135.037\,n_l^2-2.35778\, n_l^3\,,
\label{eq::deltae2}
\\[5mm]
{\delta^{(2)}_e}(3)&=&{15697\over 4}-{18215\over 24}n_l
+{1687\over 36}n_l^2-{25\over 27}n_l^3\,,
\nonumber\\
{\delta^{(1)}_e}(3)&=&{188921\over 32}+{72947\over 324}\pi^2
+{8349\over 2}\zeta(3)-{99\over 16}\pi^4
+\left(-{362177\over 288}-{4583\over 162}\pi^2-{2354\over 3}\zeta(3)
\right.
\nonumber\\
&+&\left.{3\over 8}\pi^4\right)n_l+\left({67909\over 864}
+{11\over 9}\pi^2+{422\over 9}\zeta(3)\right)n_l^2
+\left(-{13\over 9}-{2\over 81}\pi^2-{8\over 9}\zeta(3)\right)n_l^3\,,
\nonumber\\
{\delta^{(0)}_e}(3)&=&16157.3-3111.10\,n_l+185.835\,n_l^2-3.30878\, n_l^3\,,
\label{eq::deltae3}
\\[5mm]
{\delta^{(2)}_e}(4)&=&{35387\over 8}-{20393\over 24}n_l
+{1885\over 36}n_l^2-{28\over 27}n_l^3\,,
\nonumber\\
{\delta^{(1)}_e}(4)&=&{2871661\over 432}+{47671\over 216}\pi^2
+{11011\over 2}\zeta(3)-{99\over 16}\pi^4
+\left(-{1222979\over 864}-{9005\over 324}\pi^2-{3080\over 3}\zeta(3)
\right.
\nonumber\\
&+&\left.{3\over 8}\pi^4\right)n_l+\left({229655\over 2592}
+{11\over 9}\pi^2+{554\over 9}\zeta(3)\right)n_l^2
+\left(-{4771\over 2916}-{2\over 81}\pi^2-{32\over 27}\zeta(3)\right)n_l^3\,,
\nonumber\\
{\delta^{(0)}_e}(4)&=&19849.6-3844.67\,n_l+230.754\,n_l^2-4.15765\, n_l^3\,,
\label{eq::deltae4}
\\[5mm]
{\delta^{(2)}_e}(5)&=&{192907\over 40}-{110677\over 120}n_l
+{10217\over 180}n_l^2-{152\over 135}n_l^3\,,
\nonumber\\
{\delta^{(1)}_e}(5)&=&{305326847\over 43200}+{117653\over 540}\pi^2
+{13673\over 2}\zeta(3)-{99\over 16}\pi^4
+\left(-{8154553\over 5400}-{111311\over 4050}\pi^2
\right.
\nonumber\\
&-&\left.{3806\over 3}\zeta(3)+{3\over 8}\pi^4\right)n_l+\left({6135349\over 64800}
+{11\over 9}\pi^2+{686\over 9}\zeta(3)\right)n_l^2
+\left(-{255247\over 145800}-{2\over 81}\pi^2\right.
\nonumber\\
&-&\left.{40\over 27}\zeta(3)\right)n_l^3\,,
\nonumber\\
{\delta^{(0)}_e}(5)&=&23217.6-4508.70\,n_l+271.381\,n_l^2-4.93013\, n_l^3\,,
\label{eq::deltae5}
\end{eqnarray}
\begin{eqnarray}
{\delta^{(2)}_e}(6)&=&{206217\over 40}-{117937\over 120}n_l
+{10877\over 180}n_l^2-{6\over 5}n_l^3\,,
\nonumber\\
{\delta^{(1)}_e}(6)&=&{261142397\over 36000}+{69961\over 324}\pi^2
+{16335\over 2}\zeta(3)-{99\over 16}\pi^4
+\left(-{56127979\over 36000}-{13255\over 486}\pi^2
\right.
\nonumber\\
&-&\left.{4532\over 3}\zeta(3)+{3\over 8}\pi^4\right)n_l+\left({10577453\over 108000}
+{11\over 9}\pi^2+{818\over 9}\zeta(3)\right)n_l^2
+\left(-{109931\over 60750}-{2\over 81}\pi^2\right.
\nonumber\\
&-&\left.{16\over 9}\zeta(3)\right)n_l^3\,,
\nonumber\\
{\delta^{(0)}_e}(6)&=&26327.3-5118.61\,n_l+308.686\,n_l^2-5.64252\, n_l^3\,.
\label{eq::deltae6}
\end{eqnarray}
The third-order coefficients of the series for the $n=2,\ldots,6$ leptonic width read
\begin{eqnarray}
{\delta^{(2)}_z}(2)&=&{73821\over 8}-{55007\over 54}\pi^2+\left(-{88763\over 48}
+{15775\over 81}\pi^2\right)n_l+\left({8555\over 72}
-{110\over 9}\pi^2\right)n_l^2
\nonumber\\
&+&\left(-{133\over 54}+{20\over 81}\pi^2\right)n_l^3\,,
\nonumber\\
{{\delta'}^{(1)}_z}(2)&=&-{30799\over 27}-{550241\over 4860}\pi^2
-{2750\over 9}\zeta(3)+{1540\over 81}\pi^4
-{2284\over 81}\pi^2\ln{2}
\nonumber\\
&+&\left({13012\over 81}+{935\over 162}\pi^2+{500\over 27}\zeta(3)
-{280\over 243}\pi^4+{56\over 27}\pi^2\ln{2}\right)n_l
+\left(-{428\over 81}+{32\over 81}\pi^2\right)n_l^2\,,
\nonumber\\
{\delta^{(1)}_z}(2)&=&{2037613\over 144}-{94029647\over 38880}\pi^2
+{393635\over 48}\zeta(3)+{640915\over 10368}\pi^4
-{26\over 27}\pi^2\ln{2}
\nonumber\\
&+&\left(-{5585245\over 1728}+{2200685\over 3888}\pi^2-{28705\over 18}\zeta(3)
-{232505\over 15552}\pi^4+{28\over 9}\pi^2\ln{2}\right)n_l
+\left({128953\over 576}\right.
\nonumber\\
&-&\left.{25475\over 648}\pi^2+{1165\over 12}\zeta(3)+{55\over 54}\pi^4\right)n_l^2
+\left(-{257\over 54}+{799\over 972}\pi^2-{50\over 27}\zeta(3)
-{5\over 243}\pi^4\right)n_l^3\,,
\nonumber\\
{\delta^{(0)}_z}(2)&=&-4893(4)+410.99(2)\,n_l-2.04062\,n_l^2+0.0372517\, n_l^3\,,
\label{eq::deltaz2}
\\[5mm]
{\delta^{(2)}_z}(3)&=&{60203\over 4}-{169909\over 108}\pi^2+\left(-{139583\over 48}
+{47885\over 162}\pi^2\right)n_l+\left({13175\over 72}
-{55\over 3}\pi^2\right)n_l^2
\nonumber\\
&+&\left(-{203\over 54}+{10\over 27}\pi^2\right)n_l^3
\nonumber\\
{{\delta'}^{(1)}_z}(3)&=&-{46045\over 27}-{2755463\over 14580}\pi^2
-{2750\over 9}\zeta(3)+{770\over 27}\pi^4
-{2284\over 27}\pi^2\ln{2}\,,
\nonumber\\
&+&\left({18556\over 81}+{859\over 162}\pi^2+{500\over 27}\zeta(3)
-{140\over 81}\pi^4+{56\over 27}\pi^2\ln{2}\right)n_l
+\left(-{596\over 81}+{16\over 27}\pi^2\right)n_l^2\,,
\nonumber\\
{\delta^{(1)}_z}(3)&=&{19134865\over 576}-{337523323\over 58320}\pi^2
+{593285\over 48}\zeta(3)+{580085\over 3456}\pi^4
-{26\over 27}\pi^2\ln{2}
\nonumber\\
&+&\left(-{12268147\over 1728}+{892237\over 729}\pi^2-{84635\over 36}\zeta(3)
-{183415\over 5184}\pi^4+{28\over 9}\pi^2\ln{2}\right)n_l
+\left({819079\over 1728}\right.
\nonumber\\
&-&\left.{2180\over 27}\pi^2+{1715\over 12}\zeta(3)+{55\over 24}\pi^4\right)n_l^2
+\left(-{239\over 24}+{271\over 162}\pi^2-{25\over 9}\zeta(3)
-{5\over 108}\pi^4\right)n_l^3\,,
\nonumber\\
{\delta^{(0)}_z}(3)&=&-5571(4)+430.50(2)\,n_l+0.440551\,n_l^2+0.00996215\, n_l^3\,,
\label{eq::deltaz3}
\end{eqnarray}
\begin{eqnarray}
{\delta^{(2)}_z}(4)&=&{2999183\over 144}-{57451\over 27}\pi^2+\left(-{569999\over 144}
+{32110\over 81}\pi^2\right)n_l+\left({53275\over 216}
-{220\over 9}\pi^2\right)n_l^2
\nonumber\\
&+&\left(-{1226\over 243}+{40\over 81}\pi^2\right)n_l^3\,,
\nonumber\\
{{\delta'}^{(1)}_z}(4)&=&-{20390\over 9}-{1230931\over 4860}\pi^2
-{2750\over 9}\zeta(3)+{3080\over 81}\pi^4
-{2284\over 27}\pi^2\ln{2}
\nonumber\\
&+&\left({24056\over 81}+{6977\over 1458}\pi^2+{500\over 27}\zeta(3)
-{560\over 243}\pi^4+{56\over 27}\pi^2\ln{2}\right)n_l
+\left(-{2288\over 243}+{64\over 81}\pi^2\right)n_l^2\,,
\nonumber\\
{\delta^{(1)}_z}(4)&=&{16780655\over 288}-{397851877\over 38880}\pi^2
+{792935\over 48}\zeta(3)+{3318755\over 10368}\pi^4
-{26\over 27}\pi^2\ln{2}
\nonumber\\
&+&\left(-{62469101\over 5184}+{24249269\over 11664}\pi^2-{27965\over 9}\zeta(3)
-{998665\over 15552}\pi^4+{28\over 9}\pi^2\ln{2}\right)n_l
\nonumber\\
&+&\left({12280831\over 15552}-{86975\over 648}\pi^2+{755\over 4}\zeta(3)+{110\over 27}\pi^4\right)n_l^2
+\left(-{12014\over 729}+{8069\over 2916}\pi^2\right.
\nonumber\\
&-&\left.{100\over 27}\zeta(3)
-{20\over 243}\pi^4\right)n_l^3\,,
\nonumber\\
{\delta^{(0)}_z}(4)&=&-5970(4)+416.97(2)\,n_l+3.85644\,n_l^2-0.0754395\, n_l^3\,,
\label{eq::deltaz4}
\\[5mm]
{\delta^{(2)}_z}(5)&=&{7650137\over 288}-{289699\over 108}\pi^2+\left(-{45010\over 9}
+{80555\over 162}\pi^2\right)n_l+\left({33463\over 108}
-{275\over 9}\pi^2\right)n_l^2
\nonumber\\
&+&\left(-{6145\over 972}+{50\over 81}\pi^2\right)n_l^3\,,
\nonumber\\
{{\delta'}^{(1)}_z}(5)&=&-{761861\over 270}-{631136\over 2025}\pi^2
-{2750\over 9}\zeta(3)+{3850\over 81}\pi^4
-{2284\over 27}\pi^2\ln{2}
\nonumber\\
&+&\left({1822\over 5}+{3080\over 729}\pi^2+{500\over 27}\zeta(3)
-{700\over 243}\pi^4+{56\over 27}\pi^2\ln{2}\right)n_l
+\left(-{13922\over 1215}+{80\over 81}\pi^2\right)n_l^2\,,
\nonumber\\
{\delta^{(1)}_z}(5)&=&{227792547\over 2560}-{4074337157\over 259200}\pi^2
+{992585\over 48}\zeta(3)+{5376415\over 10368}\pi^4
-{26\over 27}\pi^2\ln{2}
\nonumber\\
&+&\left(-{1869252041\over 103680}+{728770499\over 233280}\pi^2-{139085\over 36}\zeta(3)
-{1577765\over 15552}\pi^4+{28\over 9}\pi^2\ln{2}\right)n_l
\nonumber\\
&+&\left({181570829\over 155520}-{172165\over 864}\pi^2
+{2815\over 12}\zeta(3)+{1375\over 216}\pi^4\right)n_l^2
+\left(-{5658707\over 233280}+{47765\over 11664}\pi^2\right.
\nonumber\\
&-&\left.{125\over 27}\zeta(3)
-{125\over 972}\pi^4\right)n_l^3\,,
\nonumber\\
{\delta^{(0)}_z}(5)&=&-6216(4)+386.70(2)\,n_l+7.76141\,n_l^2-0.151118\, n_l^3\,,
\label{eq::deltaz5}
\\[5mm]
{\delta^{(2)}_z}(6)&=&{1935901\over 60}-{174797\over 54}\pi^2+\left(-{1449083\over 240}
+{48445\over 81}\pi^2\right)n_l+\left({14907\over 40}
-{110\over 3}\pi^2\right)n_l^2
\nonumber\\
&+&\left(-{6149\over 810}+{20\over 27}\pi^2\right)n_l^3\,,
\nonumber\\
{{\delta'}^{(1)}_z}(6)&=&-{2277877\over 675}-{5343263\over 14580}\pi^2
-{2750\over 9}\zeta(3)+{1540\over 27}\pi^4
-{2284\over 27}\pi^2\ln{2}
\nonumber
\end{eqnarray}
\begin{eqnarray}
&+&\left({873628\over 2025}+{8821\over 2430}\pi^2+{500\over 27}\zeta(3)
-{280\over 81}\pi^4+{56\over 27}\pi^2\ln{2}\right)n_l
+\left(-{27316\over 2025}+{32\over 27}\pi^2\right)n_l^2\,,
\nonumber\\
{\delta^{(1)}_z}(6)&=&{7209136189\over 57600}-{1296002359\over 58320}\pi^2
+{1192235 \over 48}\zeta(3)+{2637745\over 3456}\pi^4
-{26\over 27}\pi^2\ln{2}
\nonumber\\
&+&\left(-{239985727\over 9600}+{126975799\over 29160}\pi^2-{83155\over 18}\zeta(3)
-{762515\over 5184}\pi^4+{28\over 9}\pi^2\ln{2}\right)n_l
\nonumber\\
&+&\left({2600603\over 1620}-{148811\over 540}\pi^2+{3365\over 12}\zeta(3)+{55\over 6}\pi^4\right)n_l^2
+\left(-{6462751\over 194400}+{4577\over 810}\pi^2\right.
\nonumber\\
&-&\left.{50\over 9}\zeta(3)
-{5\over 27}\pi^4\right)n_l^3\,,
\nonumber\\
{\delta^{(0)}_z}(6)&=&-6365(4)+346.96(1)\,n_l+11.9236\,n_l^2-0.232450\, n_l^3\,
\label{eq::deltaz6}
\end{eqnarray}


\end{document}